\documentclass[1p]{elsarticle} 

\usepackage{lineno,hyperref}
\usepackage[utf8]{inputenc}
\usepackage{float}
\usepackage[table]{xcolor}
\usepackage{enumitem}
\usepackage{pgfplots}
\usepackage{amssymb}
\usepackage{pifont}

\newcommand{\cmark}{\ding{53}}
\pgfplotsset{width=7cm,compat=1.14}
\modulolinenumbers[10]

\usepackage{tabularx}
\usepackage{booktabs}

\usepackage{longtable}
\usepackage{array}
\newcolumntype{C}[1]{>{\centering\arraybackslash}p{#1}}
\usepackage{xtab,afterpage}

\usepackage{graphicx}
\usepackage{subcaption}

\usetikzlibrary{mindmap}

\makeatletter
\newcommand{\aftertwo}[1]{\afterpage{\if@firstcolumn #1
  \else\afterpage{#1}\fi}}
\makeatother

\usepackage[resetlabels,labeled]{multibib}
\newcites{R}{Reviewed Articles}

\usepackage{balance}

\usepackage{xparse}
\let\oriCiteR\citeR

\RenewDocumentCommand{\citeR}{O{} O{} m}{%
  \renewcommand{\citenumfont}[1]{R##1}%
  \oriCiteR[#1][#2]{#3}%
  \renewcommand{\citenumfont}[1]{##1}%
}

\begin{document}

\begin{frontmatter}

\title{Smart Grids Co-Simulations: Survey \& Research Directions}


\author[First]{Peter Mihal}

\author[Second]{Martin Schvarcbacher}

\author[First]{Bruno Rossi}

\author[First]{Tomáš Pitner}

\address[First]{Faculty of Informatics, Masaryk University, Brno, Czech Republic}
\address[Second]{Faculty of Science, University of Amsterdam, Netherlands}

%
%
%


\begin{abstract}
The integration of renewable sources, communication and power networks with information and communication technologies is one of the main challenges in Smart Grids (SG) large-scale testing. For this reason, the coupling of simulators is commonly used to dynamically simulate several aspects of the SG infrastructure, in the so-called co-simulations.
In this paper, we provide a scoping review of research of co-simulations in the context of Smart Grids: i) research areas and research problems addressed by co-simulations, ii) specific co-simulation aspects focus of research, iii) typical coupling of simulators in co-simulation studies. Based on the results, we discuss research directions of future SG co-simulation research in each of the identified areas.

\end{abstract}


\end{frontmatter}

\section{Introduction}
The complexity of the modern smart power distribution network, the Smart Grid (SG), presents key challenges for the systems being developed. The plethora of technologies, communication protocols, and algorithms requires integrated approaches for system development, testing and validation~\citep{ref:buscher2015using}. 
One solution to tackle such complexity is to use simulations to test different scenarios and interactions with the environment---such as modelling renewable production and energy pricing interactions in microgrids~\citep{ref:luo2018distributed,ref:hwang2017energy}. However, isolated simulations alone are not enough to represent the dynamic behaviors and interplays between a variety of systems and the complex energy-related ecosystems: coupling of multiple simulation environments involving both software and hardware devices is necessary to create realistic scenarios~\citep{ref:besanger2017using, ref:vogt2018survey}.

In the SG context, both the power and the data communication domain can be modelled concurrently to understand complex interactions, often with some hardware-in-the-loop (HiL) support \citep{ref:li2014simulation-review}. Co-simulations have been useful for many research goals in the SG area, such as modelling power failures and recovering capability of the power network, investigating device failures, simulating electrical vehicles charging for demand / response management, simulating the impact of communication packets loss on the power network, as well as different forms of data injection attacks. Such coupling of different simulators poses several challenges in time-synchronization between the simulators, being some simulators based on discrete events, such as network simulators, while others on continuous time, such as power simulators \citep{ref:vogt2018survey}. Furthermore, such simulations have to be contextualized to SG testbeds experiments \citep{ref:cintuglu2017survey}, large infrastructures for testing of SG components in terms of both functional and non-functional requirements. To conduct such large-scale experiments, multiple locations can be used and thus be part of the experiment~\citep{ref:buscher2015using}.  Communication across large distances and multiple hardware / software solutions, can bring even more stringent real-time and synchronization challenges~\citep{ref:buscher2015using}.  

The coupling of distinct simulators, each one running in its own runtime environment, is commonly defined as \textit{co-simulation}. A co-simulator allows the connection of multiple software simulators and hardware emulators to enable multiple unified simulation scenarios~\citep{ref:vogt2018survey}. In this way, multiple domains can be tested and simulated transparently as being part of a single system. Connecting multiple simulators presents many challenges, which are addressed in different ways by the various co-simulation frameworks, such as the mentioned time-synchronization, and the type of architecture for orchestrating the simulators.

The goal of this paper is to provide a scoping review of co-simulations usage in the SG domain by giving an aggregated view about the research performed: i) research areas and research problems addressed by SG co-simulations, ii) specific SG co-simulation aspects focus of research, iii) typical coupling of simulators in SG co-simulation studies. All the knowledge from the review is then used to delineate future research directions.

The structure of the paper is as follows: Section~\ref{sec:intro-simulations-cosimulations} introduces the concept of SGs, typical research focus on co-simulations and results from previous reviews. Section~\ref{survey-sg-cosimulations} presents the methodology of conducting the review together with the review needs, and the set of research questions to answer. Section~\ref{sec:results} presents the main results by answering the main research questions together with threats to validity: what problems were addressed by co-simulations, the main focus of research, and the main simulators used. Section~\ref{sec:research_dir} presents the main research directions for all the SG areas reviewed in this article, while Section~\ref{sec:conclusion} provides the conclusions.

\section{Co-Simulations for Smart Grids}\label{sec:intro-simulations-cosimulations}

A Smart Grid is an electricity network that represents the convergence of Information and Communication Technology (ICT), sensors, and power systems to supply electricity to consumers via two-way digital communication with the goals to improve reliability, efficiency, and resilience \citep{ref:fang2011smart}. Controllers, sensors, computer systems, automation equipment work together to provide efficient transmission of electricity, fast restoration of electricity after power disturbances, reduced costs for utilities, lower power costs for consumers, reduced peak demand, increased integration of large-scale renewable energy systems, better integration of customer-owner power generation systems, and improved security. The sustainability and security of the existing communication network is reached by adding digital infrastructures to the electrical grid to achieve real-time monitoring of power generation and distribution~\citep{goel2015smart}.

The area of SGs is highly complex due to the intersection of multiple multidisciplinary areas, from the services offered, to the technical aspects and the social impacts. Over time, different conceptual reference models have been proposed to represent the SGs. The CEN-CENELEC-ETSI standardization Group used as base the same conceptual model proposed by the National Institute of Standards and Technology (NIST), to create the Smart Grid Architecture Model (SGAM) \citep{ref:bruin2012sgam}.

SGAM provides a three-dimensional model including domain, zones and layers, to represent the complex SG context \citep{ref:bruin2012sgam}. The zones constitute different levels of power system management and are constituted by process, field, station, operation, enterprise, and market levels. For example, the station level can provide the information-level abstraction for data concentration, and substation automation.
The domains represent the energy conversion flow \citep{ref:bruin2012sgam}: generation, transmission, distribution, distributed electrical resources (DER), customer premises. For example, the distribution level constitute the infrastructure for power distribution to customers.  
The layers represent different cross-cutting concerns for SG architectures, and are represented by the component, communication, information, function, and business layers. The component layer is composed of physical and virtual devices, the communication layer by the protocols used, the information layer by the information models, the function layer by the functionality requirements, and the business layer by the business goals for the provided services. The SGAM view provides a way to consider the variety of aspects involved in SG-related implementation and research.

\subsection{SG Research Areas}
\label{sec:research-areas}

So far, the research in the SG domain has covered a variety of aspects. To classify co-simulation articles in this review, we used a classification of SG research in nine main areas (A1-A9) as defined in~\citet{ref:cintuglu2017survey}, in which the last two areas (A8.~Cybersecurity and A9.~Network communications) are meant to be seen as complementary to the research goals in the other main seven areas.

\noindent \textbf{A1. Reliability and wide-area awareness}. The aim of the situational awareness research is to diagnose, anticipate, and respond to prevent problems before disruptions arise~\citep{ref:cintuglu2017survey,NIST2010}. In the context of SGs, the self-healing property means that SGs can detect issues and resume normal operations. By this automated process, the time needed to repair is significantly reduced~\citep{Amin2008,Beidou2010}. Achieving this requires an important effort put into strong and reliable protection, control and communication network~\citep{Amin2008}.

\noindent \textbf{A2. Consumer energy efficiency.} The goal is to lower energy use during times of peak demand or when the power reliability is at risk~\citep{ref:cintuglu2017survey}. This effect is crucial for optimizing the balance of the distributed power~\citep{NIST2010}.

\noindent \textbf{A3. Distributed   Energy   Resources   (DER).} This research area covers energy storage and utility-independent generation units. The generated power is mainly consumed at the prosumer premises as a negative load~\citep{ref:cintuglu2017survey,NIST2010}. The integration of renewable energy sources with the grid is also an important aspect of research. Alternative power sources such as wind or photovoltaic units provide additional power to the grid and make it more stable, especially during peak demand times. In addition, the generation of this kind of electric power provides a more environmental-friendly output~\citep{Beidou2010}.

\noindent \textbf{A4. Grid energy storage.} The research focuses on the development of new storage capabilities. The energy storage concept covers the conversion of electrical energy from a power network into a form of energy that can be stored and converted back to electrical energy~\citep{Mohd,ref:cintuglu2017survey,NIST2010}. A very promising research aspect is the usage of plug-in electric vehicles, which bring a way to store additional energy (known as \textit{vehicle-to-grid}).

\noindent \textbf{A5. Electric Transportation.} Research in this area mainly focuses on wired-wireless, battery banks, large-scale grid integration and charging stations~\citep{ref:cintuglu2017survey}. Another significant research aspect is the opportunity to use electric vehicles as a mobile power storage unit~\citep{NIST2010}.

\noindent \textbf{A6. Advanced Metering Infrastructure.} The research on the power quality side mostly focuses on smart meters technology in order to localize and detect different types of distortion. A smart meter is an electronic device that monitors and records electric energy consumption and communicates the information to the electricity supplier for monitoring and billing. 
The main research goal of this area is an integration of various technologies that provide an intelligent connection between consumers and system operators~\citep{Zhang}. System operators implement demand response and price signaling mechanism to serve according to dynamic pricing~\citep{ref:cintuglu2017survey,NIST2010}.

\noindent \textbf{A7. Management of distribution grid.} Advanced cyber-physical architectures for distribution grid management aim to maximize the performance of feeders, transformers and other components of the networked distribution systems and integrate them with transmission systems and customer operations~\citep{ref:cintuglu2017survey,NIST2010}.

\noindent \textbf{A8. Cybersecurity.} Since SGs are built on top of ICT infrastructures, they are vulnerable to cyber-attack threats. Cybersecurity in the SG context considers specific communication protocols in various domains~\citep{ref:cintuglu2017survey,NIST2010}. In addition, the electric grid is very sensitive and represents a potential national target, increasing the gains and motivations for attackers~\citep{Wang2013}. 

\noindent \textbf{A9. Network communications.} SGs make use of two-way communication to provide improved protection, monitoring, and optimization for all grids components. Customers use a variety of public and private communication networks, both wired and wireless~\citep{Zaballos}. Prosumer network communication adopts mainly Home Area Network (HAN) to intelligently manage devices. Wireless machine-to-machine (M2M) communication between smart meters eliminates human intervention necessity to operate the grid intelligently~\citep{ref:cintuglu2017survey,NIST2010}.


\subsection{Co-simulation Aspects}
\label{sec:co-sim-aspects}
In general terms, a simulation is represented by a mathematical model describing properties of the system being modelled and an independent solver that is applied to find a more or less approximate solution \citep{ref:vogt2018survey}. The model represents key characteristics of the simulated system, which can be obtained by an abstraction of the real system. Power systems can be studied under varying conditions and scenarios. One of the benefits that simulations provide is the option to control simulation time. Researchers can adjust the running of time, so simulations can run faster than real systems. This can be helpful to reduce time required to evaluate different scenarios, while granting synchronization with real-time devices, if necessary.


Co-simulations consist of multiple simulators which are coupled together and run separately. Each can cover a different subsystem or aspect of the SG, giving results that can provide better understanding of coupling effects. Co-simulations pose several challenges in terms of the run-time infrastructure adopted and the way events are synchronized. In general, simulated systems can be of two types, either discrete or continuous in nature, depending if state variables change at fixed points in time or continuously~\citep{ref:law2000simulation}. To model such systems, discrete events and continuous time based models can be used depending on the needs of the simulations (e.g., if considering the continuous nature of a phenomenon can be important based on the research goals). One of the main issues in the integration in a co-simulation is to combine continuous time models of power systems with discrete events simulations from communication networks \citep{ref:mets2014combining}.  

\begin{itemize}
    \item \textit{Discrete Events Models} --- (co)simulators that model a system as it evolves by considering variables state changes at specific points in time, where an event is an occurrence that can modify the system's state~\citep{ref:law2000simulation}.
    \item \textit{Continuous Time Models} --- (co)simulators that consider continuous change of variable states based on the flow of time. There might be some function that expresses changes of states over time and that could be potentially solved analytically. However, the complexity might lead to the usage of simulations in the first place~\citep{ref:law2000simulation}.
        \item \textit{Hybrid } --- (co)simulators that integrate both discrete events and continuous time simulators and are not limited to one of the two instances~\citep{ref:gomes2018co-sim-survey-acm}.
\end{itemize}

\noindent In this sense, an important aspect about co-simulations is the way in which different simulators are synchronized---that is how data is exchanged between simulation solvers:

\begin{itemize}
    \item \textit{Conservative synchronization} --- in this type of synchronization, \textit{"each simulator strictly processes events in a time stamp order"}~\citep{ref:cintuglu2017survey}. For instance, a dynamically defined barrier for all simulators, which only allows a next simulation iteration after all simulators have finished. It is referred also as \textit{“barrier synchronization”}.
    \item \textit{Optimistic synchronization} --- errors are detected during the simulation and different mechanisms are used to revert them. For example, a pre-defined number of events are stored and in case of an out-of-order event, the simulation is reversed to a time before this event and executed again with this event in order; hence it is also called \textit{“Time Warp”}. The name \textit{“optimistic”} assumes that there are no causality errors~\citep{ref:law2000simulation}.
\end{itemize}

\begin{table}[!htbp]
\caption{Some of the main co-simulation platforms applied to the SG area.}
\label{tbl:co-sim-platforms}
\footnotesize
\begin{tabular}{p{0.45\linewidth}p{0.10\linewidth}p{0.20\linewidth}p{0.15\linewidth}}
\toprule
\textbf{Name} &  \textbf{Year}&  \textbf{Syncronization} &  \textbf{Architecture}\\ 
\midrule
Daccosim-NG~\citep{ref:evora2019daccosim} & 2019 & Discrete Events & FMI-based\\
CyDER~\citep{ref:nouidui2019cyder} & 2019 & Discrete Events & FMI-based\\
HELICS~\citep{ref:palmintier2017design} & 2017 & Discrete Events & Federated\\
MECSYCO~\citep{ref:camus2016mecsyco} & 2015 & Discrete Events & Ad-hoc\\
Daccosim~\citep{ref:galtier2015fmi} & 2015 & Discrete Events & FMI-based\\
FNCS~\citep{ref:ciraci2014fncs} & 2014 & Discrete Events & HLA\\
INSPIRE~\citep{ref:georg2013inspire} & 2013 & Discrete Events & HLA\\
GECO~\citep{ref:lin2012geco} & 2012 & Discrete Events & Ad-hoc \\
MOSAIK~\citep{HLAarch2} & 2011 & Discrete Events & Ad-hoc\\
VPNET~\citep{ref:li2011vpnet} & 2011 & Continuous Time & Ad-hoc\\
EPOCHS~\citep{ref:hopkinson2006epochs} & 2006 & Continuous Time & Ad-hoc \\
\bottomrule
\end{tabular}
\end{table}

\noindent The runtime infrastructure represents the mechanisms and architectures used to coordinate the different simulators within a co-simulation context~\citep{ref:cintuglu2017survey,ref:law2000simulation}. \textit{Single simulation} architectures only use one solver. In this context, communication between different simulators is not an issue. \textit{Parallel simulation} architectures are tightly coupled systems which often share the same memory and are able to perform inter-process communication. Their communication latency must be reduced to minimum to avoid the introduction of bottlenecks. \textit{Distributed simulations} architectures are more complex. often composed of several computers distributed over different remote locations with higher latency than in the parallel simulation architectures.
Methods of events synchronization can also vary in each co-simulation platform:



Co-simulation frameworks are responsible of data exchange and synchronization between different simulators. From the architecture point of view, they can follow several styles for structuring components. One way is to have a central orchestrator component that deals with the synchronization issues, but other ways of management are possible, such as federated models typical of High Level Architecture (HLA)~\citep{HLAarch}.
HLA is a domain-independent reference architecture and a standard aimed at the integration of different simulators and at the synchronization among them~\citep{HLAarch}. In 2000, it became the official IEEE-1516 standard~\citep{IEEEHLA3}.  


\begin{figure*}[!htb]
{\centering
  \makebox[0pt]{%
    \usetikzlibrary{mindmap}
  \tikzset{concept/.append style={fill={none}}}
    \begin{tikzpicture}
[mindmap, transform shape, grow cyclic, scale=0.6, text=black, every node/.style=concept, concept color=gray!140,
level 1/.append style={level distance=7cm},
level 2/.append style={level distance=4cm}]
\node{Smart Grid Co-Simulations}
child [concept color=gray!120, grow=-60, text width=2.5cm] { node {Environment benefits}
	child [grow=80]{ node {Electricity generation and distribution}}
	child [grow=50]{ node {Energy usage optimized}}
	child [grow=160]{ node {Cost reducing effects}}
	child [grow=10]{ node {Decrease of pollution levels}}
	child [grow=190]{ node {Decrease peak demand}}
}
child [concept color=gray!120, grow=240] {node {Simulation architectures}
    child [grow=-190, text width=2.5cm]{node {Heterogeneous domains}}
    child [grow=-230]{node {Single simulation architectures}}
    child [grow=-150]{node {Parallel simulation architectures}}
    child [grow=90]{node {Distributed simulations architectures}}
}
child [concept color=gray!120, grow=110]{node {Platform selection}
    child [grow=220, text width=2cm]{node {Simulation model validation}}
    child [grow=50, text width=2cm]{node {Simulation techniques support}}
    child [grow=85]{node {HiL support}}
    child [grow=120]{node {Real-time needs}}
    child [grow=150]{node {Open source support}}
    child [grow=180]{node {Scenarios supported}}
}
child [concept color=gray!120, grow=180, text width=3.7cm] { node {Applications}
	child [grow=210]{ node {Electrical supply}}
	child [grow=-120]{ node {Renewable power sources}}
	child [grow=110, text width=2.5cm]{ node {SG infrastructure optimization}}
	child [grow=150, text width=2cm]{ node {Vehicle-to-grid}}
	child [grow=180]{ node {Security modelling}}
}
child [concept color=gray!120, grow=50, text width=2.9cm] { node {Synchronization}
	child [grow=105, text width=2cm]{ node {Conservative sync. }}
	child [grow=60, text width=2cm]{ node {Optimistic sync.}}
	child [grow=0, text width=2cm]{ node {Discrete event sync.}}
	child [grow=-55, text width=2cm]{ node {Time stepped sync.}}
	child [grow=-90, text width=2cm]{ node {Barrier sync.}}
};
\end{tikzpicture}}\par}
\caption{Smart Grids simulations concept map} \label{fig:Mindmap}
\end{figure*}
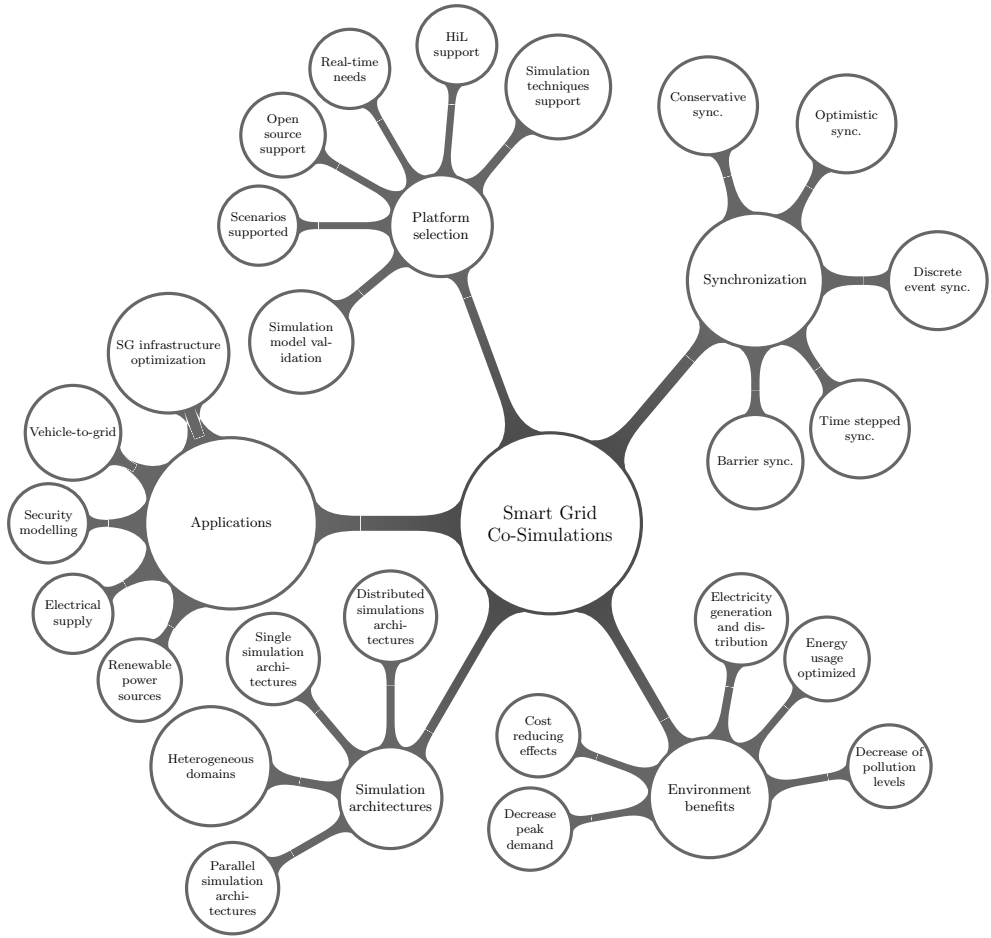

In HLA, all components of this architecture are federated and work independently. Each component is connected to a bus known as a run-time infrastructure. The run-time infrastructure bus provides services, which are responsible for data flow coordination between them (synchronization) in order to guard the correct time advancement~\citep{HLAarch,IEEEHLA3}. \citet{HLAarch} showcases the orchestration process of the OMNET++ network simulator, the Jade framework for a multi-agent system, and the Simulink modeller coordinated into a HLA architecture.

The Functional Mock-up Interface (FMI) was introduced as a standard to allow the integration and coupling of several models from different domains (e.g., mechanical, electrical)~\citep{ref:blochwitz2011functional}. Through the usage of the API provided by the standard, co-simulation platforms supporting FMI can be more easily integrate components (like the co-simulations platforms Daccosim-NG~\citep{ref:evora2019daccosim} and CyDER~\citep{ref:nouidui2019cyder}).

Over time, many co-simulation platforms have been proposed to solve mainly the issue of synchronization between simulators, starting from the EPOCHS framework that was proposed in 2006~\citep{ref:hopkinson2006epochs} to the Hierarchical Engine for Large-scale Infrastructure Co-Simulation (HELICS)~\citep{ref:palmintier2017design} and Daccosim-NG~\citep{ref:evora2019daccosim} that are the most recent frameworks proposed to address scalability and usability issues of previous co-simulators. While HLA provides a domain-independent co-simulation reference architecture that can be used, other types of \textit{"ad-hoc"} architectures  emerged over time. For example, the Framework for Network Co-simulation (FNCS) provides intentionally a lightweight set of functionalities for data exchange adopting some ideas from HLA, but not not utilizing the whole standard~\citep{ref:ciraci2014fncs}. As well, the Mosaik framework~\cite{HLAarch2}, was specifically focused on SGs, thus adopting a simpler architecture than HLA~\cite{104-Steinbrink2018-Smartgridco}.
We summarize in table \ref{tbl:co-sim-platforms} some of the main co-simulation architectures used in SG studies so far, with the year of appearance of the platform, the type of synchronization (if discrete events or continuous time), and the type of architecture (if adopting HLA or some \textit{ad-hoc} architectural style).

In this introductory part, we only scratched the complexity of co-simulations discussing aspects useful for the scoping review. Given the complexity of the domain, the interested reader can find more challenges and research problems in a recent extensive survey by~\citet{ref:gomes2018co-sim-survey-acm}, specifically focused on co-simulations. As a summary, we provide a concept map of common concepts found in the SG co-simulation domain (Fig.~\ref{fig:Mindmap}), covering several relevant aspects: applications, platform selections, synchronization aspects, architectures, and benefits.

\subsection{Previous SG co-simulation reviews}
There are not many previous surveys that focus on SG co-simulations---the main representatives being \cite{ref:vogt2018survey, ref:li2014cosim-platforms-review,ref:mets2014combining,ref:yi2016overview,ref:schloegl2015towards}. Furthermore, the focus of previous surveys is mainly on discussing and categorizing the (co)simulation platforms and characteristics, rather than looking at the application scenarios like the current survey.

\citet{ref:mets2014combining} provided an extensive overview of the area of co-simulations for power systems, focusing on aspects such as synchronization, and providing a classification of both power and network simulators typically used in the area. Authors divide the frameworks available for research in three categories: power systems (e.g., OpenDSS), communication networks (e.g., OMNeT++), and SGs simulators (e.g., GridLAB-D). Overall, the review covers twelve power system simulators, four network simulators and several SGs simulators.

\citet{ref:yi2016overview} classify co-simulation methods in three types: i) unified, ii) non-real-time, and iii) real-time simulation methods. Authors review several power and communication co-simulation platforms based on the combination of distinct frameworks, such as OpenDSS, Modelica, ns2, OMNeT++, together with the application area of the co-simulation solution (e.g., for SCADA security or wide-area monitor/control and protection). They further discuss  implications about co-simulation platform architectures for time synchronization.

\citet{ref:li2014simulation-review} propose an overview of simulation techniques available in the area of SG co-simulations, focusing more at the communication level. Authors provide a comparison of the main communication simulators used in the area (ns-2, OPNET,OMNeT++). A discussion of existing SG simulators is presented (SmartGridLab, GridSim, SCORE, GridLAB-D). Furthermore, extensions necessary to power simulators to include network simulations (and vice-versa) are discussed. Several platforms for co-simulations integrating both power and communication aspects are discussed, such as EPOCHS, GECO, and VPNET.

\citet{ref:schloegl2015towards} propose a classification scheme for co-simulation tools in SG to help developers and users to have a common understanding of the domain. Co-simulation frameworks are categorized according to time-resolution (steady-state, electro-mechanic range, electro-magnetic range), synchronization (continuous, fixed step, variable step, event driven), time ratio (faster / same / slower than wall clock time). Furthermore, authors distinguish between hybrid simulations (one solver per multiple models), co-simulations (several solvers interacting), hardware supported simulation, hardware in the loop simulation \citep{ref:schloegl2015towards}. 


\citet{ref:vogt2018survey} survey 26 SG simulation platforms (e.g., Mosaik, EPOCHS), looking at their main differences.  The final result is a classification of co-simulation platforms by research areas using the European Technology SG categorization, with indications of types of co-simulation and simulators used, correlation between research topics and simulation tools, research areas, synchronization types and real-time and HiL support. Based on the review, future research directions for co-simulations are discussed, in terms of investigation of flexibility of the markets, simulating interactions between grid operators, and to estimate cost savings during power grids expansions.

\begin{table*}[!htb]
\renewcommand{\arraystretch}{1.2}
\caption{Comparison of current review and Vogt \textit{et al.} \citep{ref:vogt2018survey}.}
\label{tbl:comparison}
\footnotesize
\begin{tabular}{p{0.22\linewidth}p{0.35\linewidth}p{0.35\linewidth}}
\toprule
\textbf{Aspect} &  \textbf{Vogt \textit{et al.} \citep{ref:vogt2018survey}}&  \textbf{This survey}\\ 
\midrule
\textbf{\# Studies} & 50 articles (26 platforms)  & 82 articles \\
\textbf{Main Focus} & SG co-sim \underline{platforms}  & SG co-sim \underline{applications} \\
\textbf{Research Method} & scoping / mapping study  & scoping / mapping study \\
\textbf{SG research areas} & 30 Research Areas from the European Technology Platform SG  & 9 Research Areas for SG testbeds defined in \cite{ref:cintuglu2017survey} \\

\textbf{Research Questions} & 
\begin{itemize}
    \item correlation btw simulation tools and research topics
    \item distribution of research areas of the co-sim platforms
    \item number of buses and distribution of simulation time spans
    \item synchronization, open source, real-time and HiL support
\end{itemize} & 

\begin{itemize}
    \item identification of research areas of application of SG co-sim.
    \item focus of the co-sim application (e.g., for power and comm integration) 
    \item simulators applied the studies and their coupling
\end{itemize} 
  \\

\bottomrule
\end{tabular}
\end{table*}

Among the reported research, the only previous study directly comparable to the current review is the survey by \citet{ref:vogt2018survey}. However, our review is focused on research articles reporting the \textit{application} of co-simulations to the context of SGs rather on the usage of co-simulations \textit{platforms} themselves. Our goal is more focused on the aspects investigated by means of co-simulation platforms, the type of components and simulators involved, and research directions for co-simulation research.
As such, the current review can complement the findings derived from~\citet{ref:vogt2018survey} that is more focused on the specific SG co-simulation platforms. We highlight the main differences between the two reviews in Table \ref{tbl:comparison}. Although not explicitly mentioned in~\citet{ref:vogt2018survey}, we can consider both survey to follow the same research methodology as a scoping study~\citep{ref:arksey2005-scoping} / mapping study~\citep{ref:petersen2008systematic,ref:barn2017conductingsms}---attempting to map the identified frameworks / studies / research outcomes to different facets to derive research gaps and future research directions.

\section{Smart Grids Co-simulations Review}\label{survey-sg-cosimulations}

Based on the published research in the area of co-simulations, we run a formal literature review about the application of co-simulations in the area of SG, in terms of focus of research, technologies and frameworks adopted. We follow the search protocol of systematic mapping studies \citep{ref:petersen2008systematic,ref:barn2017conductingsms}, in which the focus is more on collecting quantitative information about published research rather than in-depth discussion of each source. The advantage of such methodology is that it allows to get an overview of the whole research area.
We set three main research questions to drive the whole research analysis process.

\subsection{Main Research Questions}

\begin{enumerate}
    \item[RQ1.] What are the \underline{research areas} and \underline{research problems} that co-simulations studies addressed so far in the SG domain (e.g., cyber-security by simulating data injection attacks);
    \item[RQ2.] What are the \underline{specific aspects} of co-simulations in the SG domain that are the focus of the articles (e.g., it could be related to the definition of a new synchronization method);
    \item[RQ3.] What is the typical \underline{coupling of simulators} adopted in each of the case studies in the SG domain (e.g., we might find a specific power simulator, PyPower, used more often in combination with \textit{ns-3} as a network simulator);
\end{enumerate}

\noindent Based on the results answering these main research questions, we elaborate further on research applying co-simulations to the SG context, providing a series of research directions (Section \ref{sec:research_dir}).

 \subsection{Review Process}

For the review process, we selected four main digital repositories: IEEE Xplore, ACM Digital Library (DL), Springerlink, and ScienceDirect. These repositories were selected based on the heterogeneity of results that would be expected, reaching a low number of duplications.  Overall, we used the queries listed in Table~\ref{tbl:queries}.

\begin{table}[!htbp]
\caption{Review queries and total number of papers.}
\label{tbl:queries}
\footnotesize
\begin{tabular}{p{0.18\linewidth}p{0.55\linewidth}p{0.15\linewidth}}
\toprule
\textbf{Repo} &  \textbf{Query}&  \textbf{\#}\\ 
\midrule
IEEE Xplore & \texttt{Metadata (abstract+title text+indexing terms:((smart grid*) AND co-simulation*)} & 196 \\
ACM DL & (\texttt{(+smart +grid* +co-simulation*) (ANY FIELD: title, abstract, full text)} & 266 \\
SpringerLink &  \texttt{('smart AND grid AND co-simulation') (Full-text search)}  & 200\\
ScienceDirect & \texttt{("smart grid" OR "smart grids") AND ("co-simulation" OR "co-simulations") (Full-text search)} & 129\\
\bottomrule
\end{tabular}
\end{table}




 We did not set any \textit{a-priori} range for years of the queries. The overall main idea of using co-simulations in the area of SG emerged in year 2006 (see the EPOCHS framework~\cite{ref:hopkinson2006epochs}), but the work of Godfrey \textit{et al.}~[R28] was one of the first to provide some case study). 
Conversely, the general concept of co-simulations in other domains dates back longer time before (see recent co-simulation reviews by \citet{ref:gomes2017co-sim-survey,ref:gomes2018co-sim-survey-acm}, e.g. for the automotive domain usage of co-simulations dates back to 1998). 

Overall, after running the queries on the four digital repositories, we had 791 articles after the first phase of querying. By merging all the results there were 47 duplicates, so we had 744 articles in total for the scoping review. Based on the research questions of the review, we set some inclusion criteria: \textit{IC1)} papers in which one/more co-simulation frameworks were discussed in the context of SGs, \textit{IC2)} papers in which there was at least one practical example / case study of the application of the adopted co-simulation framework for a concrete SG research problem. Only articles in English were included. A first phase of filtering was done based on the title and abstract of the papers. 567 articles were removed mainly because the main focus was not about co-simulations. At this stage, we had 177 papers (56 articles to be included, and 121 to be reviewed again). The final phase of inclusion for \textit{"undecided"} papers, yielded the final 123 articles that were included in the pre-final step of the review.

The last step was about extracting all the required information from the papers to answer the four main research questions. During this process, as a collateral effect some papers were removed as not found relevant according to the goals of this review (e.g., papers not covering the definition of co-simulation frameworks, rather then just discussion about qualities of frameworks). Overall, 41 papers were removed in this step, leaving a final set of 82 papers.

\aftertwo{
\onecolumn

\scriptsize

{\setlength{\tabcolsep}{6pt}
\begin{longtable}{p{0.02\textwidth}p{0.35\textwidth}|C{0.15\textwidth}|C{0.015\textwidth}|C{0.015\textwidth}|p{0.015\textwidth}|p{0.015\textwidth}|p{0.015\textwidth}|p{0.015\textwidth}|p{0.015\textwidth}|p{0.015\textwidth}|p{0.015\textwidth}|} 
 \caption{(RQ1) Problems addressed by co-simulations and research areas.}\\
\toprule \\
 &  \textbf{Problems addressed} & co-sim platform & A1 & A2 & A3 & A4 & A5 & A6 & A7 & A8 & A9 \\  
\midrule\\
\nociteR{002-Albagli2016-Smartgridframework}{[R2]} & Power failures recovery & JADE & \cmark & & & & & & \cmark &   & \\
 \nociteR{016-Bottura2013-SITLHLAco}{[R13]} & Reliability of Monitoring and Control & Custom & \cmark &  & & &  & & \cmark & & \\
\nociteR{031-Garau2015-ICTreliabilitymodelling}{[R26]} & Reliability of the ICT network   & Matlab & \cmark & & & &  & & & & \\
\nociteR{034-Godfrey2010-ModelingSmartGrid}{[R28]} & Communication failures & Custom & \cmark & & & &  &  & \cmark & & \cmark \\
\nociteR{045-Kelley2015-federatedsimulationtoolkit}{[R35]} & Protection relays reliability & FNCS & \cmark & & & &  & & \cmark & & \\
\nociteR{049-Lai2014-Designcosimulation}{[R38]} & Power and communication failures & Custom & \cmark & & & &  & & & & \cmark \\
\nociteR{058-Lin2011-Powersystemcommunication}{[R47]} & Agent-based remote relay protection  & Custom & \cmark & & & & & & \cmark & & \cmark \\
\nociteR{060-Lin2012-GECOGlobalEvent}{[R46]} & Remote relay protection  & GECO & \cmark & & & & & & \cmark & & \cmark  \\
\nociteR{073-Mueller2012-Hybridsimulationpower}{[R56]} & Monitoring, protection and control & FNCS & \cmark & & & & & & & & \cmark  \\
\nociteR{083-Ravikumar2017-iPaCSintegrativepower}{[R64]} & Communication in control applications  & Custom & \cmark & & & & & & \cmark & \cmark & \cmark \\
\nociteR{086-Roche2012-FrameworkCosimulation}{[R66]} & Monitoring and control applications & Matlab + JADE & \cmark & & & & & & \cmark  & &   \\
\nociteR{090-Saxena2017-CPSACyberPhysical}{[R68]} & Bad data measurements     & Matlab + JADE & \cmark & & & & & & & \cmark &   \\
\nociteR{097-Shum2018-CoSimulationDistributed}{[R69]} & Agent-based fault location, restoration  & JADE & \cmark & & & & & & & &   \\
\nociteR{108-Sun2014-cosimulationplatform}{[R75]} & Reliability of the control strategies    & Matlab & \cmark & & & & & & & & \cmark   \\
\nociteR{117-Vogt2015-Evaluationinteractionsmultiple}{[R80]} & Operation, control strategies  & OpSim & \cmark & & & & & & & & \cmark   \\
\nociteR{119-Wang2017-ArchitectureApplicationReal}{[R81]} & Voltage control of photovoltaic stations  & Custom & \cmark & & & \cmark & & & \cmark & &   \\
\nociteR{121-Zhao2013-COsimulationplatform}{[R82]} & Cascading Failures & Custom & \cmark & & & & & & & \cmark &  \\
\nociteR{081-Pan2016-NS3MATLABco}{[R63]} & Reconfiguration after faults  & GECO &  \cmark & \cmark & & & & & & & \cmark \\
\nociteR{006-Armendariz2014-cosimulationplatform}{[R5]} & Power monitoring \& control & Custom & \cmark & \cmark & & & & \cmark & &  & \\
\nociteR{062-Makhmalbaf2014-Cosimulationdetailed}{[R50]}  & Demand and supply load balancing   & FNCS & \cmark & \cmark & \cmark & &  & & & &  \\
\nociteR{123-duan2020cybersecurity}{[R22]} & Cybersecurity Distribution Grid    & HELICS & & \cmark & & & & &  & \cmark & \cmark   \\
\nociteR{071-Mosshammer2013-Loosecouplingarchitecture}{[R54]} & Voltage control distribution power grid & Custom & \cmark & \cmark & \cmark & & & & & &  \\
\nociteR{004-Alishov2016-Cosimulationarchitecture}{[R3]} & Distribution network models & Custom &  & \cmark & & & \cmark & & \cmark &  & \\
\nociteR{005-Amarasekara2015-Cosimulationplatform}{[R4]} & Power distribution network events & Custom &  & \cmark & & & & & &  & \cmark \\
\nociteR{008-Awad2016-Cosimulationbased}{[R7]} & Voltage profiles & Sgsim &  & \cmark  & & & & & \cmark &  & \\
\nociteR{010-Ayon2017-Integrationbottomstatistical}{[R8]} & Users power consumption behaviour & Custom &  & \cmark & & &  & &  &  & \\
\nociteR{015-Bompard2016-multisitereal}{[R12]} & Control strategies for prosumers & Custom &  & \cmark & & \cmark & \cmark & & &  & \\
\nociteR{026-Duerr2017-Loadbalancingenergy}{[R23]} & Optimization techniques for SGs & EnergyPlus & & \cmark & & \cmark &  & & & & \\
\nociteR{038-Hess2016-Multivariatepowerflow}{[R31]} & Residential loads balancing & Mosaik & & \cmark & & & & & \cmark & & \\
\nociteR{051-Latif2016-Cosimulationbased}{[R39]} & Thermostatically controlled loads  & Custom &  & \cmark & & &  & & & & \\
\nociteR{053-Lehnhoff2015-Exchangeabilitypowerflow}{[R42]} & Grid modelling analysis   & Mosaik & & \cmark & \cmark & &  & & & &  \\
\nociteR{072-Moulema2015-EffectivenessSmartGrid}{[R55]} & Demand / response and energy pricing     & FNCS & & \cmark & & & & \cmark & & & \cmark  \\
\nociteR{074-Nannen2015-Lowcostintegration}{[R57]} & Integration of low-cost HiL & Mosaik & & \cmark & \cmark & & & \cmark & & &   \\
\nociteR{078-Otte2018-HardwareLoopCo}{[R60]} & Power system control applications     & Lablink & \cmark & \cmark & \cmark & & & & & &   \\
\nociteR{069-Mittal2015-SystemsystemsApproach}{[R53]} & Home energy management and tariffs   & Custom & & \cmark & \cmark & & & \cmark & & &  \\
\nociteR{023-Ciraci2014-FNCSFrameworkPower}{[R20]} & SG market applications for pricing  & FNCS & & & \cmark & &  & \cmark & & & \\
\nociteR{025-Ding2016-Investigationgriddriven}{[R21]} & Real-time market-grid coupling  & GridLAB-D & & & \cmark & &  & \cmark & & & \\
\nociteR{039-Huang2017-Opensourceframework}{[R32]} & Transmission / distribution networks & FNCS & & & \cmark & & & & \cmark & & \\
\nociteR{044-Kazmi2016-flexiblesmartgrid}{[R34]} & Voltage regulation in distribution & JADE & & & \cmark & & \cmark & & & & \\ 
\nociteR{063-Mallapuram2017-IntegratedSimulationStudy}{[R51]} & Power load and market pricing   & FNCS & & & \cmark & &  & \cmark & & &  \\
\nociteR{068-Mirtaheri2016-frameworkcontrolco}{[R52]} & Control algor. distribution network  & Custom & \cmark & & \cmark & & \cmark  & & & &  \\
\nociteR{048-Kounev2015-microgridcosimulation}{[R37]} & Power supply to offshore production & Matlab & & & \cmark & &  & & \cmark & & \\
\nociteR{047-Kosek2014-Evaluationsmartgrid}{[R36]} & Power-balancing in the isolated grid & Mosaik + JADE & \cmark & & \cmark & \cmark &  &  & & & \\
\nociteR{017-Broderick2017-Techniqueinterconnectcontrol}{[R14]} & Electric Vehicles charging events & Custom &  &  & & \cmark & \cmark & &  & & \\
\nociteR{056-Li2017-DistributedLargeScale}{[R44]} & Stable grid charging electric vehicles   & GridLab-D & & & & \cmark & \cmark  & & & & \cmark \\
\nociteR{096-Shum2013-developmentsmartgrid}{[R71]} & Vehicle to grid voltage support     & FNCS & & & & \cmark & \cmark & & & & \cmark \\
\nociteR{106-Stifter2013-Cosimulationcomponents}{[R74]} & Electric Vehicles smart charging    & Modelica & & & & \cmark & \cmark & & & &   \\
\nociteR{043-Johnstone2017-Cosimulationapproach}{[R33]} & Voltage control in generators & Matlab &  & & & & \cmark & & \cmark & & \\
\nociteR{054-Levesque2012-Communicationspowerdistribution}{[R43]} & Control algorithms of electric vehicles  & Custom & \cmark & & \cmark & \cmark & \cmark & & & &  \\
\nociteR{080-Palensky2013-ModelingIntelligentEnergy}{[R61]} & Flexible electric vehicles charging  & Modelica & & & & \cmark & \cmark & \cmark & & &   \\
\nociteR{011-Barbierato2018-DistributedIoTInfrastructure}{[R9]} & Demand Response management & Custom &  &  & & &  & \cmark &  &  & \\
\nociteR{112-Tariq2014-CyberphysicalCo}{[R77]} & Communication Demand Response  & Custom & & & & & & \cmark & & & \cmark  \\
\nociteR{110-Syed2015-Ancillaryserviceprovision}{[R76]} & Demand side management for services & Custom & & & & & & \cmark & & &   \\
 \nociteR{030-Fuller2013-Communicationsimulationspower}{[R24]} & Pricing, and comm. delays in DR & GridLab-D & & & & &  & \cmark & & & \cmark \\
\nociteR{018-Bytschkow2015-CombiningSCADACIM}{[R15]} & Synchronization for SCADA & EPOCHS & &  & & &  & \cmark & \cmark & & \\
\nociteR{115-Troiano2016-Cosimulatorpower}{[R78]} & Power load balancing  & Custom & & & & & & & \cmark & & \cmark  \\
\nociteR{105-Stevic2013-twostepsimulation}{[R73]} & Power distribution voltage control  & Custom & & & & & & & \cmark & &   \\
\nociteR{013-Bhor2016-Networkpowergrid}{[R10]} & Wide-area grid monitoring & Custom & \cmark  & \cmark  & &  & & & \cmark &  &  \\
\nociteR{021-Celli2016-CosimulationICT}{[R18]} & Network voltage regulation  & Matlab & & & & &  & & \cmark & & \cmark \\
\nociteR{001-Ahmad2015-Cosimulationframework}{[R1]} & Voltage regulation power distribution & JADE & & & & & & & \cmark &   & \\
\nociteR{050-Latif2015-alternatePowerFactoryMatlab}{[R40]} & Voltage control & Matlab & & & & &  & & \cmark & & \\
\nociteR{075-Nguyen2017-Usingpowerhardware}{[R58]} & Power and ICT testing & Mosaik & & & & & & & \cmark & & \\
\nociteR{059-Lin2012-Cybersecurityimpacts}{[R46]} & Injected data and cyber attacks   & Custom & & & & & & & \cmark & \cmark & \cmark   \\
\nociteR{037-Hammad2019-Implementationdevelopmentoffline}{[R30]} & Control sys resilience to cyber threats & Matlab & & & & & & & & \cmark & \\
\nociteR{099-Stefanov2012-ICTmodelingintegrated}{[R72]} & Simulating cyberattacks    & Matlab & & & & & & & \cmark & \cmark &   \\

\nociteR{019-Caire2013-Vulnerabilityanalysiscoupled}{[R16]} & Grids vulnerabilities identification & Custom & & & & &  & & \cmark & \cmark & \\
\nociteR{020-Cao2018-SimplifiedCoSimulation}{[R17]} & Data attacks to the grid  & Custom & \cmark & & & &  & & & \cmark & \cmark \\
\nociteR{014-Bian2015-Realtimeco}{[R11]} & Distribution comm. performance & Custom &  &  & & &  & & \cmark &  & \cmark \\
\nociteR{022-Chromik2017-Contextawarelocal}{[R19]} & Context-aware intrusion detection  & Mosaik & & & & &  & \cmark & & \cmark & \\
\nociteR{032-Garau2017-EvaluationSmartGrid}{[R25]} & Communication in energy distribution & Matlab & & & & &  & & \cmark & & \cmark \\
\nociteR{033-Georg2013-INSPIREIntegratedco}{[R27]} & Network communication monitoring & Custom & \cmark & & & &  & & & & \cmark \\
\nociteR{035-Gurusinghe2016-CosimulationPower}{[R29]} & Communication network performance  & Custom & & & & &  &  & \cmark & & \cmark \\

\nociteR{052-Lau2012-developmentsmartgrid}{[R41]} & Grid components communication & Custom & \cmark & & & &  & & & & \cmark \\
\nociteR{057-Liberatore2011-Smartgridcommunication}{[R45]} & Voltage control  real-time comm.   & Modelica & \cmark & & & & & & & \cmark  & \cmark \\
\nociteR{061-Liu2018-RealTimeCo}{[R49]} & HiL cyber-attacks  & Custom & & & & &  & & \cmark & \cmark & \cmark  \\
\nociteR{076-Ni2018-cyberphysicalpower}{[R59]} & Cyber-attacks on voltage control     & Matlab & & & & & & & \cmark & \cmark &   \\
\nociteR{082-Pan2017-Cosimulationcyber}{[R62]} & Data attacks in energy management     & Matlab &  & & & & & & \cmark & \cmark &   \\
\nociteR{084-Razaq2015-Simulatingsmartgrid}{[R65]} & Network simulators power distribution  & GridLab-D & & & & & & & \cmark & & \cmark   \\
\nociteR{088-Sadi2015-CoSimulationPlatform}{[R67]} & SCADA systems cyber-attacks & Matlab & \cmark & & & & & & & \cmark &   \\
\nociteR{091-Schloegl2016-PerformancetestingSmart}{[R69]} & Data transmission voltage control     & Mosaik & & & & & & & \cmark & & \cmark \\
\nociteR{007-Awad2014-SGsimsimulationframework}{[R6]} &  Electrical vehicle recharging & Sgsim & \cmark &  & \cmark & \cmark &  & &  &  & \cmark \\
\nociteR{116-Venkataramanan2016-Realtimeco}{[R79]} & Data communication attacks microgrid    & Custom & & & & & & & \cmark & \cmark & \cmark   \\
\bottomrule
 
 \label{tbl:co-sim-problems}


\end{longtable}
}
}

\section{Main Review Results}\label{sec:results}






\subsection{Research Areas and Research Problems (RQ1)}
We first address the research areas and problems that co-simulations are meant to address in the papers (e.g., cyber-security by simulating data injection attacks). For this goal, we mapped all the co-simulation articles in the categories of SGs research~(A1-A9) defined in \citet{ref:cintuglu2017survey}, that we defined in Section \ref{sec:research-areas} \textit{"SG Research Areas"} (Table \ref{tbl:co-sim-problems}). 

Each paper was mapped to one or more categories, depending on the problem addressed by the usage of co-simulations: A1. Reliability and wide-area awareness~(32 papers), A2. Customer energy efficiency~(18), A3. Energy resource distribution~(16), A4. Grid energy storage~(11), A5. Electric Transportation~(11), A6. Advanced Metering Infrastructure~(14), A7. Management of distribution grid~(34),~A8. Cybersecurity~(15), A9. Network communications~(29).
Overall, co-simulations have been useful for a variety of research goals in the SG area, by allowing the coupling of different simulators, mainly power and communication ones. 

\noindent \textbf{A1. Reliability and wide-area awareness}. Co-simulations have been often used for performance and reliability of the power networks wide area monitoring and control systems using data provided by phase measurement units, generally analyzing the combined effects of communication, network, power levels for wide-area monitoring, protection and control~\citepR{016-Bottura2013-SITLHLAco,058-Lin2011-Powersystemcommunication,060-Lin2012-GECOGlobalEvent,073-Mueller2012-Hybridsimulationpower}. There are also studies that looked into bad data measurements and the impact of cyberattacks, looking at the latency and bandwith for control and protection applications, and simulating cascading failures in SGs~\citepR{083-Ravikumar2017-iPaCSintegrativepower,090-Saxena2017-CPSACyberPhysical,121-Zhao2013-COsimulationplatform}.

\noindent \textbf{A2. Customer energy efficiency}. Co-simulations have been used for load control, to model customers and prosumers behaviour, voltage control for demand and supply load balancing, simulating residential loads, studying demand and response and energy pricing \citepR{008-Awad2016-Cosimulationbased,010-Ayon2017-Integrationbottomstatistical,015-Bompard2016-multisitereal,026-Duerr2017-Loadbalancingenergy,038-Hess2016-Multivariatepowerflow,053-Lehnhoff2015-Exchangeabilitypowerflow,062-Makhmalbaf2014-Cosimulationdetailed,072-Moulema2015-EffectivenessSmartGrid}.

\noindent \textbf{A3. Energy resource distribution}. Co-simulations have been used for studying the integration of PV panels for household power consumption, optimal strategies for electrical vehicles recharging, the plug-in strategies of electric vehicles to stabilize grid voltage \citepR{074-Nannen2015-Lowcostintegration,056-Li2017-DistributedLargeScale,017-Broderick2017-Techniqueinterconnectcontrol}.

\noindent \textbf{A4. Grid energy storage}. Co-simulations have been used for electric vehicles charging events simulations in the context of the grid,  to investigate wireless and sensor technologies for electric vehicles charging, investigating flexible charging algorithms \citepR{017-Broderick2017-Techniqueinterconnectcontrol,056-Li2017-DistributedLargeScale,096-Shum2013-developmentsmartgrid,106-Stifter2013-Cosimulationcomponents,080-Palensky2013-ModelingIntelligentEnergy}.

\noindent \textbf{A5. Electric Transportation}. Co-simulation was used for simulating control algorithms for energy distribution networks, using electric vehicles as power storage units to stabilize the grid \citepR{056-Li2017-DistributedLargeScale,096-Shum2013-developmentsmartgrid,011-Barbierato2018-DistributedIoTInfrastructure}.

\noindent \textbf{A6. Advanced Metering Infrastructure}. Co-simulations were used for modelling demand / response scenarios and energy pricing, studying both the influence of physical aspects of the grid and the energy market \citepR{069-Mittal2015-SystemsystemsApproach,023-Ciraci2014-FNCSFrameworkPower,025-Ding2016-Investigationgriddriven,063-Mallapuram2017-IntegratedSimulationStudy,112-Tariq2014-CyberphysicalCo,110-Syed2015-Ancillaryserviceprovision,030-Fuller2013-Communicationsimulationspower}.

\noindent \textbf{A7. Management of the distribution grid}. Co-simulations were focused on synchronization mechanisms for SCADA systems, power feeders load balancing and power control, voltage regulation in power distribution networks, phasor measurement units regulation \citepR{018-Bytschkow2015-CombiningSCADACIM,115-Troiano2016-Cosimulatorpower,105-Stevic2013-twostepsimulation,013-Bhor2016-Networkpowergrid,001-Ahmad2015-Cosimulationframework,075-Nguyen2017-Usingpowerhardware}.
 
\noindent \textbf{A8. Cybersecurity}. Co-simulations have looked into the integration of network failures simulations, data injection attacks in different parts of the SG infrastructure, simulation of cyberattacks, evaluation of intrusion detection algorithms, identification of the vulnerabilities of the integration of the power network and ICT \citepR{099-Stefanov2012-ICTmodelingintegrated,019-Caire2013-Vulnerabilityanalysiscoupled,022-Chromik2017-Contextawarelocal,061-Liu2018-RealTimeCo,076-Ni2018-cyberphysicalpower,082-Pan2017-Cosimulationcyber,088-Sadi2015-CoSimulationPlatform,116-Venkataramanan2016-Realtimeco}. 

\noindent \textbf{A9. Network communications}. Co-simulations were deployed to analyze communication networks performance in the integration within the SG infrastructure, real-time communication, evaluation of wireless and wired communication networks, communication between grid components and control systems \citepR{014-Bian2015-Realtimeco,032-Garau2017-EvaluationSmartGrid,033-Georg2013-INSPIREIntegratedco,035-Gurusinghe2016-CosimulationPower,052-Lau2012-developmentsmartgrid,057-Liberatore2011-Smartgridcommunication,091-Schloegl2016-PerformancetestingSmart,007-Awad2014-SGsimsimulationframework}.

\subsection{Specific research aspects of SG co-simulations (RQ2)}
In this research question, we look at the main focus of the application of a co-simulation platform, in terms of the problems the solution addresses. We identified four main research focuses:

\begin{enumerate}
    \item [{F}1.] \textbf{Approaches for time and objects synchronization in co-simulations.} The focus of the co-simulation platform is about addressing the issues of time and object synchronization between two different simulation environments---e.g.~\citeR{005-Amarasekara2015-Cosimulationplatform,033-Georg2013-INSPIREIntegratedco}. For example, a synchronization method for coordinating simulators that can be tuned according to the simulation applications was presented in~\citeR{081-Pan2016-NS3MATLABco}. A novel co-simulation scheduler taking into account events from power and communication network simulators, with the timing of each embedded controller’s execution loop was proposed for synchronization in~\citeR{048-Kounev2015-microgridcosimulation}.
  
    \item[{F}2.] \textbf{Real-time monitoring and testing.} The focus is on the real-time properties of the co-simulation platform. For example, the co-simulation platform is focused on  allowing to perform real-time monitoring and control tests and simulations for MV/LV grids~\citepR{006-Armendariz2014-cosimulationplatform}. A real-time software-in-the-loop set-up which emulates the behavior of the real-world systems integrating inputs from IoT devices from customers to retrieve energy information is presented in~\citeR{011-Barbierato2018-DistributedIoTInfrastructure}.
    
    \item[{F}3.] \textbf{Integration of power and communication simulators.} Many papers discuss the effectiveness of co-simulations in analysing the coupled effects between power system and communication infrastructure, being very convenient to co-simulate different aspects instead of building a single tool with all the capabilities (power system modelling, intelligent control and communication)~\citepR{001-Ahmad2015-Cosimulationframework}. For example, a simplified co-simulation model is used to analyze the interdependencies between energy and information flows~\citepR{020-Cao2018-SimplifiedCoSimulation}. The FNCS framework using a federated co-simulation model for integrating transmission and distribution network simulators was discussed in~\citeR{039-Huang2017-Opensourceframework}.
    
    \item[{F}4.] \textbf{Co-simulation architecture.} The focus is on the evaluation of different components within a co-simulation platform. For example, several aspects of co-simulations, required to develop a framework, together with the simulation architecture design are discussed in \citeR{002-Albagli2016-Smartgridframework}. The architecture and configuration of two different co-simulation approaches, SITL (System in the Loop) and HLA are discussed in~\citeR{016-Bottura2013-SITLHLAco}. Comparison of running simulations with Mosaik support are discussed in~\citeR{038-Hess2016-Multivariatepowerflow,053-Lehnhoff2015-Exchangeabilitypowerflow}. Loose coupling of heterogeneous components (i.e. continuous and time-triggered subsystems) is evaluated by means of a message bus, to allow multiple simulators to exchange messages in~\citeR{071-Mosshammer2013-Loosecouplingarchitecture}.
    
\end{enumerate}

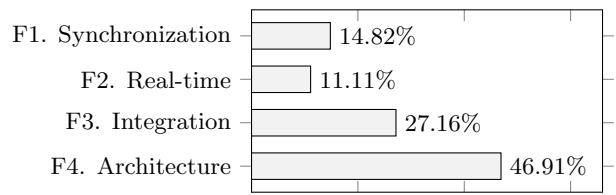
\begin{figure}[!htb]
\centering
\begin{tikzpicture}
\small
 \begin{axis}[
 		xbar, 
 		xmin=0,
 		xmax=66,
        width=6.2cm, height=4cm, enlarge y limits=0.2, 
        symbolic y coords={F1. Synchronization, F2. Real-time, F3. Integration, F4. Architecture},
 		ytick=data,
 		xticklabel=\empty,
 		y dir=reverse,
 		point meta={x},
 		nodes near coords={\pgfmathprintnumber\pgfplotspointmeta\%},
        nodes near coords align={horizontal}
 ]
 		\addplot+[fill=gray!10,draw=black,text=black] coordinates {(14.82,F1. Synchronization) (11.11,F2. Real-time) (27.16,F3. Integration) (46.91,F4. Architecture)};
 \end{axis}
\end{tikzpicture}
\caption{Focuses of co-simulation research}
\label{fig:focus-co-sim}
\end{figure}

We mapped all papers to each of the four main types of focus of the articles. Looking at the number of papers (Fig. \ref{fig:focus-co-sim}), the definition of co-simulation platforms was focused, in increasing order of frequency, on F4. Architecture~(46.91\%), F3. Integration~(27.16\%), F1. Synchronization aspects~(14.82\%), and F2. Real-time aspects~(11.11\%). While running a complete trend analysis would be inconclusive due to the sample size and differences in each category, the impression is that the discussion about architectural aspects was more the focus of recent years of publications.

\begin{figure}[!htb]
\centering
\includegraphics[width=0.97\linewidth]{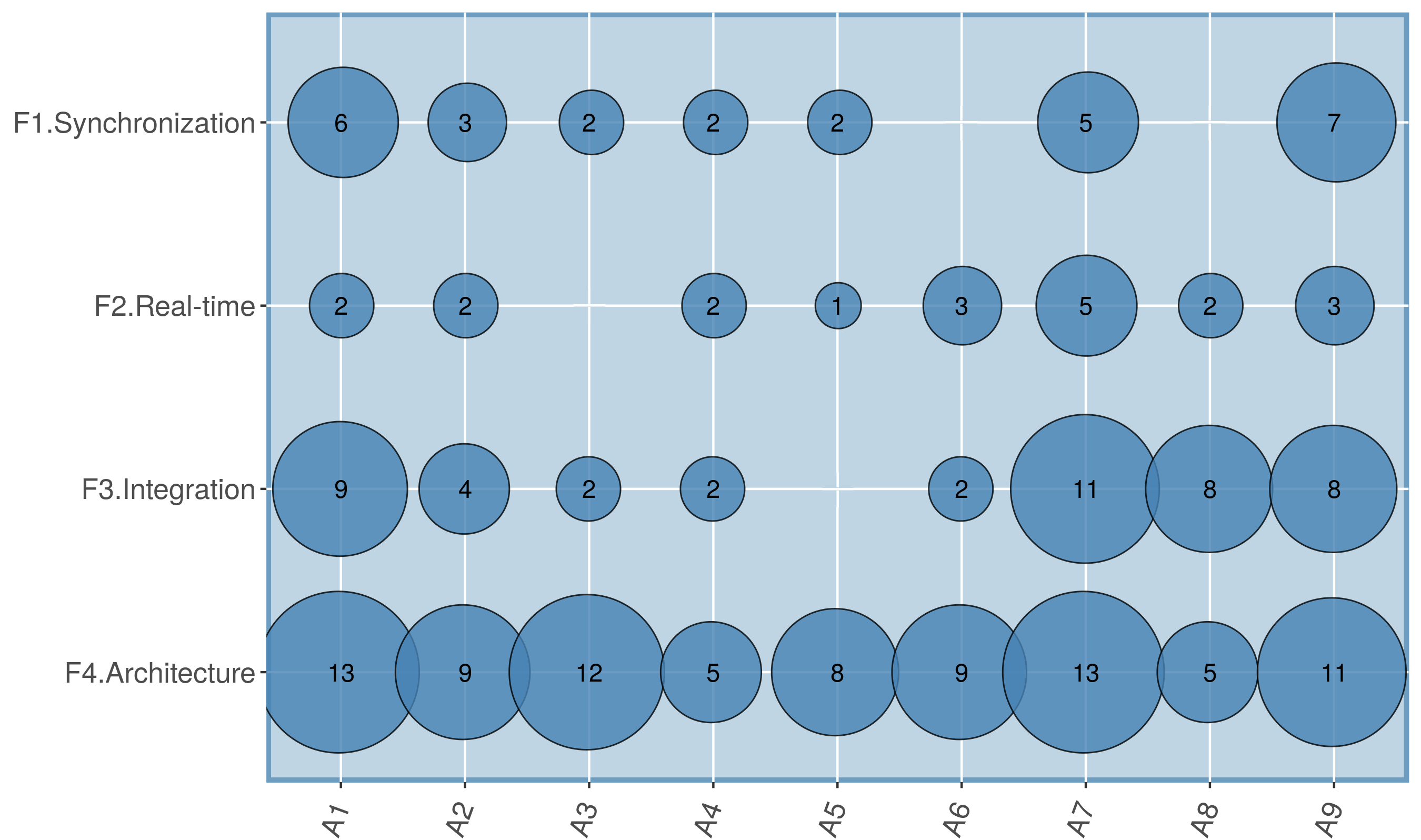}
\caption{F1-F4 co-simulation focuses by SG research area}
\label{fig:focus-co-sim-F1-F4-A1-A9}
\end{figure}

Considering the SG research areas (A1-A9) previously discussed, we then mapped each type of focus by the area of research in Fig. \ref{fig:focus-co-sim-F1-F4-A1-A9}. This view can give a representation of the distribution of the research interest by each of the identified categories.

We can see that \textit{F1.Synchronization} aspects were more investigated in papers related to the categories \textit{A1. Reliability and wide-are awareness} and \textit{A9. Network communications}. \textit{F2.Real-time} aspects were more related to \textit{A6. Advanced Metering Infrastructure}, \textit{F3.Integration} aspects more for \textit{A8. Cybersecurity}, \textit{A9. Network communications}, and \textit{F4. Architecture} aspects more for \textit{A2. Customer energy efficiency}, \textit{A7. Management of distribution grid}, and \textit{A9. Network communications}.

The adoption of co-simulations faces several challenges from the integration of platform from different domain, and scaling to a level that can be considered comparable to real-world electricity grids.

About the issues in the creation of co-simulation platforms, FMI (Section \ref{sec:co-sim-aspects}) was introduced as a way to reduce the complexity of coupling simulators from different domains via a common API~\citep{ref:blochwitz2011functional}. The standard has been effectively adopted by platforms such as Daccosim-NG~\citep{ref:evora2019daccosim} and CyDER~\citep{ref:nouidui2019cyder}. However, the standard was not considered for platforms such as HELICS~\citep{ref:palmintier2017design} due to scalability concerns over a certain number of federates and for high-speed requirements.

While HLA (Section \ref{sec:co-sim-aspects}) aims to support multiple development environments and platforms~\citep{HLAarch,IEEEHLA3}, some co-simulation platforms take a different approach. For example, the Mosaik framework~\cite{HLAarch2}, differently from HLA, was built as platform for co-simulations specifically focused on SGs. Being focused on the context, the architecture was simplified, based on a simulation manager and a scheduler ~\cite{104-Steinbrink2018-Smartgridco}. Conversely, the HELICS platform considered the runtime infrastructure in HLA of open source implementations not to be scalable over 100K federates. However, HLA principles and time synchronization approaches were considered for the design of HELICS~\citep{ref:palmintier2017design}.

\subsection{Coupling of simulators adopted (RQ3)}

In this research question, we looked into the specific simulators that are used: the main simulators mentioned are mostly power, communication, and general purpose simulators~(Table \ref{tbl:main-sim}).

\begin{table*}[!htb]
\caption{Main simulators and frequency of usage in the surveyed articles.}
\label{tbl:main-sim}
\footnotesize
\begin{tabular}{p{0.13\textwidth}p{0.10\textwidth}p{0.20\textwidth}p{0.35\textwidth}p{0.08\textwidth}}
\toprule
\textbf{Simulator} & \textbf{Type} &  \textbf{URL}&  \textbf{Description}  &  \textbf{Freq}\\ 
\midrule
GridLab-D & Power & \tiny{\url{https://www.gridlabd.org}} &  Power distribution system simulation and analysis tool & 16\\
JADE & Agent-based & \tiny{\url{https://jade.tilab.com}} &  Java-based Open Source platform for agent-based applications & 6  \\
Matlab Simulink  & General & \tiny{\url{https://www.mathworks.com/products/simulink.html}} &  Design and simulation software & 6  \\
MATPOWER & Power & \tiny{\url{http://www.pserc.cornell.edu/matpower/}} &  Power system simulation and optimization software & 3  \\
NS-2 & Network & \tiny{\url{https://www.isi.edu/nsnam/ns/}} &  Network communication simulator & 8  \\
NS-3 &  Network & \tiny{\url{https://www.nsnam.org}} &  Network communication simulator & 12  \\
Opal-RT & Power & \tiny{\url{https://www.opal-rt.com/software-rt-lab/}} &  Real-time simulation software & 5  \\
OMNeT++ & Network & \tiny{\url{https://omnetpp.org}} &  Network communication simulator & 14  \\
OpenDSS & Power & \tiny{\url{https://www.epri.com/#/pages/sa/opendss}} &  Power distribution system simulator & 11 \\
OpenModelica & General & \tiny{\url{https://openmodelica.org}} &  Open source modeling and simulation environment based on the Modelica language & 7 \\
OPNET & Network & \tiny{\url{https://www.riverbed.com/gb/products/steelcentral/opnet.html}} &  Network communication simulator & 10  \\
PowerFactory & Power & \tiny{\url{https://www.digsilent.de/en/powerfactory.html}} &   Power system analysis software & 13\\
PowerWorld & Power & \tiny{\url{https://www.powerworld.com/}} &  Power system simulator & 3  \\
PSCAD & Power & \tiny{\url{https://hvdc.ca/pscad/}} &  Power system simulator & 6  \\
PSLF & Power & \tiny{\url{https://www.geenergyconsulting.com/practice-area/software-products/pslf}} &  Power system analysis software & 3  \\
PYPOWER & Power & \tiny{\url{https://pypi.org/project/PYPOWER/}} &  Power system analysis software, port of MatPower to Python & 3  \\
\bottomrule
\end{tabular}
\end{table*}

As a first step, we looked at the frequency of usage of the main simulators mentioned (Table \ref{tbl:main-sim}). GridLab-D (16) was the most used power simulator, followed by PowerFactory (13), and OpenDSS (11). However, compared to the other simulators, GridLab-D provides a whole management environment that can be also used for coordination of simulators, so its high usage level is justified by the more functionality offered compared to other frameworks providing only power simulation functionality. In fact, Mets \textit{et al.} \citep{ref:mets2014combining} consider it as a whole SG simulator, providing more functionality than a simple power simulator. For networks simulators, OmNeT++ (14), NS-3 (11), OPNET (10), and NS-2 (8)  were the most used ones. There is, however, a good level of variability of the adoption of the simulators mentioned. There are also some temporal variations in the usage of the simulators, like NS-2 that is mostly mentioned in articles from years 2010-2011, less used in recent years due to the newer version NS-3.

\begin{figure}[!htb]
\centering
\includegraphics[width=0.95\linewidth]{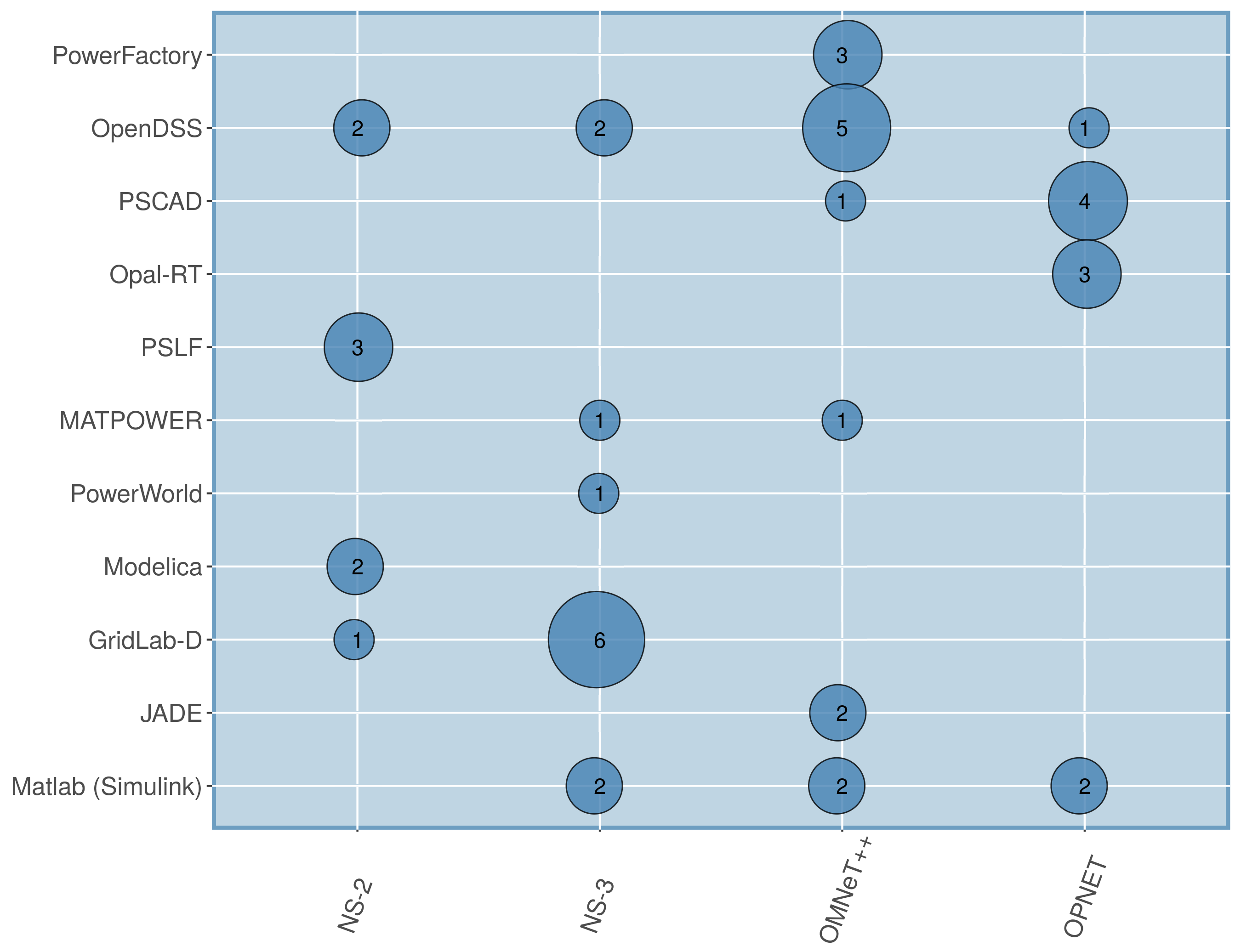}
\caption{Simulators used in combination with network simulators}
\label{fig:heat-sim}
\end{figure}

As a second step, we looked at how are the simulators combined for co-simulation purposes (Fig. \ref{fig:heat-sim}).  We divided the simulators in three categories: power, network, and others. In the third category we  included frameworks that can be used for different purposes, like GridLab-D that can be used as power simulator or for coordination of simulators, as previously mentioned. For Modelica, Matlab/Simulink, and JADE, these are used to either model or implement different aspects related to simulations. 

While we observed that OPNET is the most used network simulator in the studies provided (even though NS-3, NS-2, and OMNeT++ follow close), we can see in the heatmap that OPNET is mostly used with PSCAD and Opal-RT. OMNeT++ is more used in combination with OpenDSS, PowerFactory and JADE. NS-2 is more applied together with Modelica, OpenDSS, and PSLF, while NS-3 is more used coupled with GridLab-D, with OpenDSS, MatPower, PowerWorld and Matlab/Simulink that follow. These couplings represent the cases in which integration between the different simulators can be considered more easier from the implementation point of view.

Furthermore, we looked into the adoption of three different architecture-related concepts in the SG domain: i) the usage of agent-based systems within the architecture, ii) the support of the HLA, and iii) the support of HiL devices. Agent-based systems are an architectural style that is used in the context of simulations. The main idea is to model the behaviour of several software agents and collecting the outcomes from the evolving interactions, where an agent is an autonomous entity that can interact with the simulation environment. The capabilities of learning and adapting are key characteristics of the software agents \citep{ref:law2000simulation}. HiL represents the availability of hardware devices that can be integrated by means of the co-simulation platform. Either interfaces or other means of support need to be present in the co-simulation platform. As described in Section \ref{sec:co-sim-aspects}, HLA provides a reference architecture for integrating different simulators, so it is interesting to know the level of usage in the SG domain.

\begin{figure}[!htb]
\centering
\includegraphics[width=0.99\linewidth]{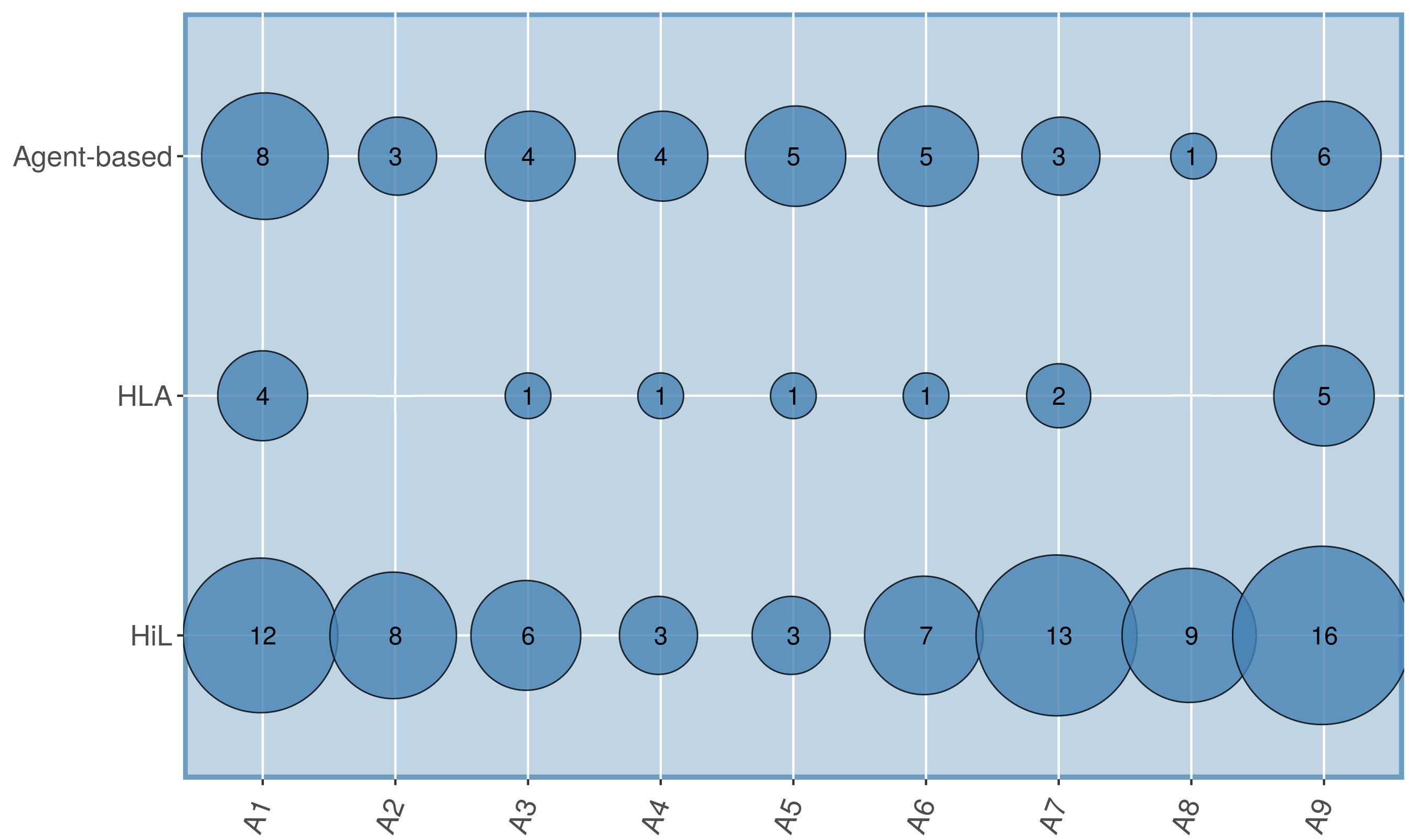}
\caption{A1-A9 categories mapped to agent-based, HAL and HiL support.}
\label{fig:map-A1-A9-AB-HAL-HIL}
\end{figure}


The findings about the distribution of these three aspects (agent-based, Hil, HLA) are quite compelling (Fig. \ref{fig:map-A1-A9-AB-HAL-HIL}). Agent-based systems are quit often used in the different SG research areas, with some (such as A4-A5 in which almost half of the articles report the usage of some form of agent-based system---typically adopting the JADE framework.
HiL is available and discussed in many of the platforms, with the studies proposing the integration with hardware devices in a large number of cases (like 70\% of the cases in category A7). Integration of hardware devices is thus an important aspect in the context of SG research.
HLA was not instead widely discussed, one main reason being that many co-simulation platforms (see Section \ref{sec:co-sim-aspects}, Table \ref{tbl:main-sim}) are using some ad-hoc architecture, while only federated solutions discuss the application of HLA standard.





\section{Research Directions}
\label{sec:research_dir}

The review confirmed some trends of research that were reported in previous studies (e.g., from \citet{ref:steinbrink2018future}), like the fact that the discussion between discrete / continuous time events was one aspect mostly focused in earlier studies, while nowadays a common focus seems more on the scalability of co-simulation studies. While in the reviewed studies we could not find yet a discussion about big data analysis, it is one research direction that will acquire more importance in the next years, together with ways to scale the analysis to larger number of sensors and IoT devices.

From the point of view of the aspects investigated, a trend we observed is that the initial focus of co-simulation papers was more on pure energy-related issues (such as simulations for voltage control), while nowadays quality aspects such as network communication, security, and privacy play a major role. We expect this trend to become even more significant in future years. The expectations are that co-simulations studies will acquire more importance for these aspects, for example for integrating better cyber-security threats analysis into the context of the power network. In this view, the coupling between power and communication simulators seems an acquired reality that is used in the majority of the case studies to represent the complexity of the SG infrastructure.

Another trend we expect to be increasing in the next years is the inclusion of HiL devices. In our review, we found that 59\% of the articles had some form of interaction with hardware devices. We expect that with the explosion of the IoT movement and the large availability of commodity hardware such as Arduinos and Raspberry PIs, such trend will increase, allowing hybrid simulations to take place in this context, as recent research has already shown---\cite{094-Schvarcbacher2017-SmartGridsCo,ref:schvar2018SGTMP}, \citeR{074-Nannen2015-Lowcostintegration}.

Another trend we expect to see is an increase in the discussion of co-simulation in the context of SG testbeds (e.g. \cite{ref:cintuglu2017survey}), leading to needs in terms of the integration in larger-scale contexts of the research performed so far in the area. This can mean more needs of integration of distributed hardware devices and software simulators. While remote access requirements have been limited so far, we expect such needs to increase in the future, leading to similar systems as DeterLab \citep{ref:mirkovic2012teaching} for teaching cybersecurity testing and simulation scenarios. Integration of the connection between cloud-based testing and expensive laboratory devices for remote real-time testing can be an alternative over cheaper solutions based on commodity IoT devices for the emulation of real hardware---like showed in \cite{094-Schvarcbacher2017-SmartGridsCo,ref:aurilio2014low}, \citeR{074-Nannen2015-Lowcostintegration}. There are many challenges involved in such "scaling-up", but also many opportunities in terms of the complexity of the simulated scenarios, sharing of knowledge between researchers and educational opportunities.

More fine-grained research directions can be based on the categorization of relevant SG research into different research areas that we used to map co-simulation research (A1-A9 categories)~\cite{ref:cintuglu2017survey}.

\noindent \textbf{A1. Reliability and wide-area awareness}. Reliable protection, control and communication networks will continue to be a focus of research supported by co-simulation frameworks. We expect that self-healing mechanisms will play an increasing role in the future, by means of simulations about interruption of services, and the evaluation of the efficiency of self-healing mechanisms. Integration of hardware devices and software simulators can help in reaching these goals.


\noindent \textbf{A2. Consumer energy efficiency.} This was an area in which earlier co-simulation studies were focused and will still be relevant area of research. We expect co-simulations to still continue to help for the evaluation of new algorithms for customers behaviour prediction and optimization, aiding in combining power simulators and real-time power systems. We expect also the increase of usage of IoT devices and commodity hardware to emulate hardware components such as smart meters.

\noindent \textbf{A3. Distributed   Energy   Resources   (DER).} We expect the area to continue being focus of research for alternative power sources such as wind or photovoltaic units to provide additional power to the grid and make it more stable, especially during peak demand times. Co-simulations can help in identifying optimizations for load balancing of the overall network.


\noindent \textbf{A4. Grid energy storage.} We expect this area to grow in terms of research related to plug-in electric vehicles, that can bring additional ways to store energy and create vehicles-to-grid networks. In the review we had already several articles discussing integration of electric vehicles, we expect more frameworks to emerge to simulate different aspects. One example is \citetR{015-Bompard2016-multisitereal} in which authors are testing a management strategy for distributed storage and 120 vehicles-to-grid, connected to a real distribution network model.


\noindent \textbf{A5. Electric Transportation.} We expect co-simulations to play an important role in this area, considering the growing needs of energy transportation, e.g. thinking about charging stations and battery banks. Also in this context, the usage of electrical vehicles as mobile power storage units will make use of simulations relevant, in similar way as the area \textit{A4. Grid energy storage}. As such, we expect more research in the line of  \citetR{080-Palensky2013-ModelingIntelligentEnergy}, looking to create a versatile platform for simulating
electric vehicle charging algorithms for demand response, coupling a Modelica-based physical simulation engine, a power network simulation tool and an agent-based simulator.



\noindent \textbf{A6. Advanced Metering Infrastructure.}  Research about smart metering devices played a large role in initial SG co-simulation research. The possibility to emulate / simulate large number smart metering devices allowed to look at the scalability of the solutions provided. In this area, the usage of co-simulation platforms with power and communication simulators has reached a certain maturity, and will still continue to be relevant for the provision of SG services.

\noindent \textbf{A7. Management of distribution grid.} This area will continue to be supported by co-simulations for the optimization of power distribution systems. In this area, we see even more the interest of the integration of renewable energy power sources within the electric grid that can benefit from co-simulation and HiL research. These, along with a forecast-based production and efficient energy storage systems still need to be investigated~\citep{Beidou2010}.

\noindent \textbf{A8. Cybersecurity.} While initial research on co-simulation was more focused on power and communication networks, we expect the area of cyber-security to become more relevant for co-simulations. We expect larger scale integration with SG testbeds~\cite{ref:cintuglu2017survey}, in which co-simulation can be integrated with security scenarios defined in cyber-security ranges~\citep{ref:vykopal2017kypo}. In any case, we expect these aspects to be more and more integrated into SG co-simulation scenarios.

\noindent \textbf{A9. Network communications.}  Initial co-simulation frameworks were focused on the integration of power and communication networks for simulating packet transmission. We expect in this area a constant move towards more complex scenarios, e.g. the best wireless-wired scenarios to connect smart meters and data concentrators, packet losses, and data injection attacks. We see this area more and more connected to A8. Cybersecurity, as reliability of the information flow within the SG infrastructure is a key element for the correctness of operations.  Furthermore, both simulations and HiL can be useful in this area to determine the most efficient communication means.

\section{Conclusions}
\label{sec:conclusion}

To address the complexity of the SG infrastructure, multiple simulation environments are used to capture the dynamic aspects of the interplay between the many systems, sensors, communication means, and energy-related ecosystems. Co-simulations were adopted in the SG domain as a way to couple and synchronize different simulators to grant more realistic scenarios involving also hardware devices.

The goal of this paper was to provide an aggregated view about the usage of co-simulations in the SG context by means of a scoping review: i) research areas and research problems addressed by SG co-simulations, ii) specific SG co-simulation aspects focus of research, iii) typical coupling of simulators in SG co-simulation studies. Based on the reviewed studies, we delineated future research directions for SG co-simulations.

In general, co-simulations have been used for a variety of goals and cross-cutting concerns. They have been used initially as a way to combine both power and communication aspects of the grid, but later on to integrate other views, such as market pricing simulations or vehicle-to-grid aspects that are quite relevant in recent years. Co-simulations have been useful for many research goals in the SG area, such as modelling power failures and recovering capability of the power network, investigating device failures, simulating electrical vehicles charging for demand/response management, simulating the impact of communication packets loss on the power network, as well as to investigate different forms of data injection attacks. Overall, our final remark is that co-simulations will continue to play a major role with even greater challenges to be addressed in the future, like the needs of integration in the context of the emergence of larger SG testbeds and new emerging cyber-threats that will challenge existing countermeasures. Furthermore, the increasing interest in renewables and electric vehicles integration within the grid can constitute a relevant application domain for co-simulation platforms.

\section*{Acknowledgments}
The research was supported from ERDF/ESF "CyberSecurity, CyberCrime and Critical Information Infrastructures Center of Excellence" (No. CZ.02.1.01/0.0/0.0/16\_019/0000822).

\renewcommand{\bibfont}{\scriptsize}




\renewcommand{\bibfont}{\footnotesize}
\bibliographystyle{model5-names}\biboptions{authoryear}



\renewcommand{\refname}{Reviewed Articles}

\bibliographystyleR{model5-names}\biboptions{authoryear}
 \tiny{

}

\end{document}